\def\lapp{\mathbin{\raise2pt \hbox{$<$} \hskip-9pt \lower4pt
\hbox{$\sim$}}}
\def\gapp{\mathbin{\raise2pt \hbox{$>$} \hskip-9pt \lower4pt
\hbox{$\sim$}}}

\newcommand{\beq}{\begin{equation}}
\newcommand{\eeq}{\end{equation}}
\newcommand{\beqar}{\begin{eqnarray}}
\newcommand{\eeqar}{\end{eqnarray}}
\newcommand{\beqarr}{\begin{eqnarray*}}
\newcommand{\eeqarr}{\end{eqnarray*}}
\newcommand{\barr}[1]{\begin{array}{#1}}
\newcommand{\earr}{\end{array}}

\newcommand{\bitem}{\begin{itemize}}
\newcommand{\eitem}{\end{itemize}}
\newcommand{\benum}{\begin{enumerate}}
\newcommand{\eenum}{\end{enumerate}}
\newcommand{\btab}[1]{\begin{tabular}{#1}}
\newcommand{\etab}{\end{tabular}}
\newcommand{\bcen}{\begin{center}}
\newcommand{\ecen}{\end{center}}

\def\({\left(}
\def\[{\left[}
\def\l\{{\left\{}
\def\){\right)}
\def\]{\right]}
\def\r\}{\right\}}
\def\what{\widehat}
\def\raw{\rightarrow}
\def\bA{\bar A}
\def\am{\bar A_{\rm max}}
\def\cf{{\cal F}}

\def\etal{{\sl et al.} }
\def\ratio#1#2{{{#1}\over{#2}}}

\def\DXDYCZ#1#2#3{\left({\partial#1\over\partial#2}\right)_{#3}}

\documentclass{aa}
\usepackage{graphicx}
\usepackage{amsmath}
\begin{document}

%%%   \thesaurus{02.08.1; 02.09.1; 09.10.1; 11.10.1}
%
   \title{X-ray emission from expanding cocoons}

%\author{}
   \author{C. Zanni \inst{1} \and G. Bodo\inst{2} \and P. Rossi \inst{2}  
   \and S. Massaglia\inst{1} \and A. Durbala \inst{3} \and A. Ferrari \inst{1}
            }

   \offprints{S. Massaglia}

   \institute{Dipartimento di Fisica Generale dell'Universit\`a,
Via Pietro Giuria 1, I - 10125 Torino\\
              email: username@ph.unito.it
\and
INAF -- Osservatorio Astronomico di Torino, Strada dell'Osservatorio
             20, I-10025 Pino Torinese\\
             email: username@to.astro.it
\and
Universitatea din Bucuresti, Facultatea de fizica atomica si nucleara,
Bucuresti-Magurele\\
PO-BOX MG-11, Romania
             }

   \date{Received; accepted }

\abstract{X-ray observations of extragalactic radiosources show strong evidences
of interaction between the radio emitting plasma and the X-ray
emitting ambient gas. In this paper we perform a detailed
study
of this interaction by numerical simulations. We study the propagation
of an axisymmetric supersonic jet in an isothermal King atmosphere and
we analyze the evolution of the resulting X-ray properties and their
dependence on the jet physical parameters. We show the existence of
two distinct and observationally different regimes of interaction,
with strong and weak shocks. In the first case shells of enhanced
X-ray emission are to be expected, while in the second case we expect
deficit of X-ray emission coincident with the cocoon. By a comparison
between analytical models and the results of our numerical
simulations, we discuss the dependence of the transition between these
two regimes on the jet parameters and we find that the mean controlling
quantity results to be the jet kinetic power. We then discuss how
the observed jets can be used to constrain the jet properties.
}

\maketitle

\section{Introduction}
X-ray observations of extragalactic radio sources have  revealed strong
evidences of  interaction between the radio emitting plasma and the
X-ray emitting gas in the ambient medium. The observation of
Cygnus A  with the ROSAT HRI (High Resolution Imager) by
Carilli, Perley \& Harris (1994) and with the Chandra X-ray Observatory
(Smith et al. 2002) showed  deficits of X-ray emission
in the cluster gas
spatially coincident with the radio lobes.  Observations
of the Perseus cluster by B\"ohringer et al. (1993), also with ROSAT,
and by Fabian et al. (2000) with the Chandra X-ray Observatory
showed that the cluster emitting gas  was
displaced by the radio lobes of the source NGC~1275.
A similar behavior has been observed in the Hydra A cluster, hosting the radio
source 3C~218, with the Chandra X-ray Observatory by McNamara et al. (2000)
(see also Nulsen et al. 2002),
and in Abell 2052 that shows regions devoid of X-ray emission coincident with
the radio lobes of 3C~217 (Blanton et al. 2001).
Other clear examples of the interaction between radio lobes and the surrounding
cluster gas are given by A~4059 (Heinz et al. 2002, Chandra) and  A~2199
(Owen \& Eilek 1998, ROSAT). Disturbances by a radio source are also found
in the gas halo of some giant elliptical galaxies such as M~87 (B\"ohringer et
al. 1995, ROSAT) and M~84 (Finoguenov \& Jones 2001, Chandra).

On the other hand, theoretical models predict that jets in
radiogalaxies inflate overpressured cocoons that displace and compress
the ambient gas and the effects of such interaction could indeed
expected to be the formation of cavities and shells in the X-ray
emission, as shown by observations. A simple one-dimensional model of
this interaction has been presented by Begelman \& Cioffi (1989). More
detailed and realistic models require the use of numerical simulations.
In this context, Clarke, Harris \& Carilli (1997) carried out
calculations of the jet propagation in a King atmosphere obtaining
simulated X-ray images to compare with ROSAT data on Cygnus A.  They
demonstrated that a deficit in the X-ray brightness is indeed shown in
the simulation results and found agreement between simulations and
observations for moderate Mach number of the jet ($M \gapp 4$).
A similar scenario is depicted by the numerical simulations of Rizza et al. (2000)
that showed the interaction and disruption of a jet inside a cooling
flow cluster. More recently
Reynolds, Heinz \& Begelman (2001) have pointed out that cocoons start being strongly
overpressured, but, during their evolution, their pressure decreases,
and they then become  essentially in pressure equilibrium with the
ambient or even underpressured. During this evolution, therefore, the
shock driven in the external medium is strong at the beginning and
becomes very weak at the end. Reynolds, Heinz \& Begelman (2001) call this last phase ``sonic
boom''. The need for weak shocks comes from the observations of cool rims
surrounding some of the X-ray cavities (see A~2052, Hydra A, Perseus A) that rules
out the possibility of strong shocks driven by the expanding cocoon.
To explain these observations several analytical (Churazov et al. 2000,
Soker et al. 2002) and numerical (Churazov et al. 2001, Brighenti \&
Mathews 2002, Quilis, Bower \& Balogh 2001) models of ``bubbles'' of hot plasma expanding
subsonically in the ambient medium have been studied. If these bubbles
are buoyant they can also explain the presence of deficits of X-ray
emission far from the radio lobes as observed in Perseus A.
On the other hand, Heinz, Reynolds \& Begelman (1998) and Alexander (2002) have studied self-similar
solutions of simplified analytical models of overpressured cocoons expanding in
a stratified medium in order to explain the observed features.

In this paper we analyze in detail, by using numerical
simulations, the evolution of the X-ray properties of expanding
cocoons and their dependence on the jet properties. The jets are characterized
by their Mach number and their density ratio with the ambient medium density; the
parameter plane is widely covered in order to consider a wide range of jet powers. We
confirm the results presented by Reynolds, Heinz \& Begelman (2001) on the existence of two distinct
and observationally different subsequent regimes of interaction, with strong or
weak shocks but we are able to determine how and when the transition between these two regimes
occurs, depending on the jet parameters.
These results, on the other
hand, show how the X-ray properties of cocoons could possibly be used
as diagnostic for the jet characteristics.

The plan of the paper is the following: in Section 2 we describe the
model, the basic equation and the initial conditions, in Section 3 we
discuss the X-ray morphologies resulting from the simulations, in
Section 4 we discuss the heating of the external material compressed
by the expanding cocoon and the consequent changes in the X-ray
emission properties, in Section 4 we discuss the physics of the cocoon
expansion that leads to the interpretation of the different X-ray
morphologies and in Section 5 we discuss the astrophysical relevance
of our results, finally a summary is presented in Section 6.

\section{Numerical simulations}

We solve numerically the  hydrodynamic equations for a supersonic
jet, in cylindrical (axial) symmetry in the coordinates $(r,z)$
continuously injected into a gravitationally stratified (but not
self-gravitating) medium
\beqar
\label{eq:syst}
\nonumber
\frac{\partial \rho}{\partial t}+\nabla\cdot(\rho \vec {v}) = 0 \\
\frac{\partial \vec {v}}{\partial t}+(\vec {v} \cdot \nabla)\vec{v}=-\nabla
p/\rho + \nabla \phi \\
\frac{\partial p}{\partial t}+(\vec{v}\cdot
\nabla)p-\gamma\frac{p}{\rho}\left[\frac{\partial p}{\partial
t}+(\vec{v}\cdot\nabla)\rho\right]= 0  \; ,
\nonumber
\eeqar

where the fluid variables $p$, $\rho$, $\vec{v}$ and $E$
are, as customary, pressure, density, velocity, and
thermal energy ($p / (\gamma -1)$) respectively;
$\gamma$ is the ratio of the specific heats. Radiative losses are neglected
since the estimated radiative times are much longer than the evolution
time scale.

The system of equations  (\ref{eq:syst}) has been solved
numerically employing a PPM (Piecewise Parabolic Method) hydrocode (Woodward
\& Colella 1984).
The integration domain has a size $r_{\rm domain}\times z_{\rm domain}$ where
$r_{\rm domain}= z_{\rm domain}=2.6a$, where $a$ is
the core radius, defined below, and
has been divided in $1024 \times 1024$ grid points. The axis of the jet is
along the left boundary of the domain ($r=0$), where we have imposed
symmetric boundary conditions for $p, \rho, v_z$ and  and
antisymmetric conditions for $v_r$. Reflective boundary conditions are
also imposed on the boundary of injection of the jet ($z=0$) outside its
radius in order to reproduce a bipolar flow and to avoid spurious inflow
effects.
Free outflow is set on the
remaining boundaries by imposing a null gradient for each variable ($d/dr=0$).

The undisturbed ambient medium is assumed stratified in a spherically
symmetric gravitational well, according to a classical isothermal King
model:
\beq
\rho_{\rm ext}(R)=\frac{\rho(0)}{\left[1+(R/a)^2 \right]^{3\beta/2}} \;,
\eeq
with  $R=\sqrt{r^2+z^2}$ and $\beta=0.5$. The resulting pressure stratification is
kept in equilibrium by an appropriated external gravitational potential.

A (cylindrical) jet is injected from the bottom
boundary of the integration domain, in pressure balance with the ambient. The
initial jet velocity profile has the form
\beq
v_z(r)=
\frac{\displaystyle v_{\rm j}}{\displaystyle \cosh[(20 \ r/a)^m]}
\eeq
and the corresponding density profile is
\beq
\rho(r,z)=
\frac {\displaystyle \rho_{\rm j}-\rho_{\rm ext}(R)}{\displaystyle \cosh[(w \
20 \ r/a)^n]}+\rho_{\rm ext}(R)
\eeq
with $w=0.77$, $m=8$ and $n=2m$, giving a jet radius $r_{\rm j} = a/20$.

Measuring lengths in units of the core radius $a$, velocities in units of the
adiabatic sound speed $c_{\rm se}$ in the undisturbed external medium and the density in
units of the ambient central density $\rho(0)$, our main parameters are the
Mach number $M \equiv v_{\rm j}/c_{\rm se}$ and the density ratio $\nu = \rho_{\rm j}/ 
\rho(0)$. Consistently the unit for the kinetic power is
\[
L_{\rm k} = \frac {\pi} {2} \rho(0) r_{\rm j}^2 c_{\rm se}^3 =
\]
\beq
= 1.5 \times 10^{42}
\left( \frac {n_0}  {10^{-2} \;{\rm cm}^{-3}} \right)
\left( \frac {a} {50 \;{\rm kpc}} \right)^2
\left( \frac {T} {3 \;{\rm keV}} \right)^{3/2}
{\rm erg \; s^{-1}}
\label{eq:Lk}
\eeq
and the unit of time is
\beq
t_0 = \frac {a}  {c_{\rm se}} = 4.8 \times 10^7
\left( \frac {a} {50 {\rm\; kpc}} \right)
\left( \frac {T} {3 {\rm\; keV}} \right)^{-1/2}
\hbox{\rm years}
\eeq
where $n_0$ is the central electron density, and we will measure time $\tau$ in units of
$t_0$.
Then the kinetic power of the jet expressed in unity of $L_{\rm k}$ is given by
\beq
L_{\rm j}=M^3 \nu L_{\rm k} \; .
\label{eq:L_j}
\eeq
In Table \ref {table:cases} we show the parameters $M$ and $\nu$ for the
simulations that we have performed giving also the corresponding values
of the kinetic jet power $L_{\rm j}$ in units of $L_{\rm k}$ and the final simulation
times in units of $t_0$.

\noindent
\begin{table}
\caption{Parameters of the seven simulations performed: Mach number $M$,
 density ratio $\nu=\rho_{\rm j}/\rho(0)$, kinetic jet power $L_{\rm j}/L_{\rm k}=M^3\nu$ and final time
 of the simulation $\tau_{\rm fin}=t_{\rm fin}/t_0$}

\begin{center}
\begin{tabular}{cccc}  \hline \hline
 $M $   & $\nu$ & $L_j/L_k$  & $\tau_{\rm fin}$ \\

\hline

10  &  0.1  &  $10^2$  &  1.38\\
\hline

60  &  0.001  & $2.16 \times 10^2$ & 1.22 \\
\hline

60 &   0.01 &  $2.16 \times 10^3$  & 0.49 \\
\hline

60 & 0.1  & $2.16 \times 10^4$ &  0.15\\
\hline

120  & 0.001 & $1.73 \times 10^3$  &  0.76\\
\hline

120  & 0.01  & $1.73 \times 10^4$ & 0.25\\
\hline

120 & 0.1    & $1.73 \times 10^5$ & 0.07\\

\hline

\end{tabular} 
\end{center} 
\label {table:cases}
\end{table}

\section{X-ray signatures: shell and cavity}

The general structure of the interaction between a low density jet and
the ambient medium is well known since the first simulations of Norman
et al. (1982) (see also Massaglia, Bodo \& Ferrari 1996 and Krause 2003)
). The flowing jet matter, slowed down by one or more
terminal shocks, inflates a cocoon that compresses and drives shocks
in the  surrounding external medium. The compressed ambient material
forms a shell surrounding the cocoon: the boundary between the shell
and the cocoon is marked by a contact discontinuity, while the
boundary between the shell and the undisturbed external medium is
marked by a shock. We will call ``{\it extended cocoon}'' the whole region
interested in the interaction between the jet and the ambient medium.
The ``{\it extended cocoon}'' is then formed by the cocoon proper and by the
surrounding shell. The cocoon, which is formed by
the expanded jet material forms a cavity with  very low density and
high temperature, in which the X-ray emissivity is strongly depressed.
On the other hand the external material in the shell has an enhanced
emission due to its compression. In addition, depending on the
external shock strength this material can be heated and its emission
properties can then change.

In principle, we then expect three main features in the
X-ray properties of the region of interaction between a jet and the
ambient medium:
\benum
\item a region of depressed emission coincident with the
cocoon;
\item a shell of enhanced emission;
\item  a variation in the spectral
properties of the emission by the shell.
\eenum
The actual appearance,
however, will depend on an interplay between all these effects. In
fact, we have to take into account that what we see is an integration
along the line of sight, that can go both through the cocoon and
through the shell. Therefore, it can happen that the enhanced shell
emission can compensate the deficit of emission in the cocoon and, in
these cases, in correspondence of the cocoon we can even see a flux
higher than that expected in the undisturbed case.

\begin{figure*}
\centering
  \includegraphics[width=17cm]{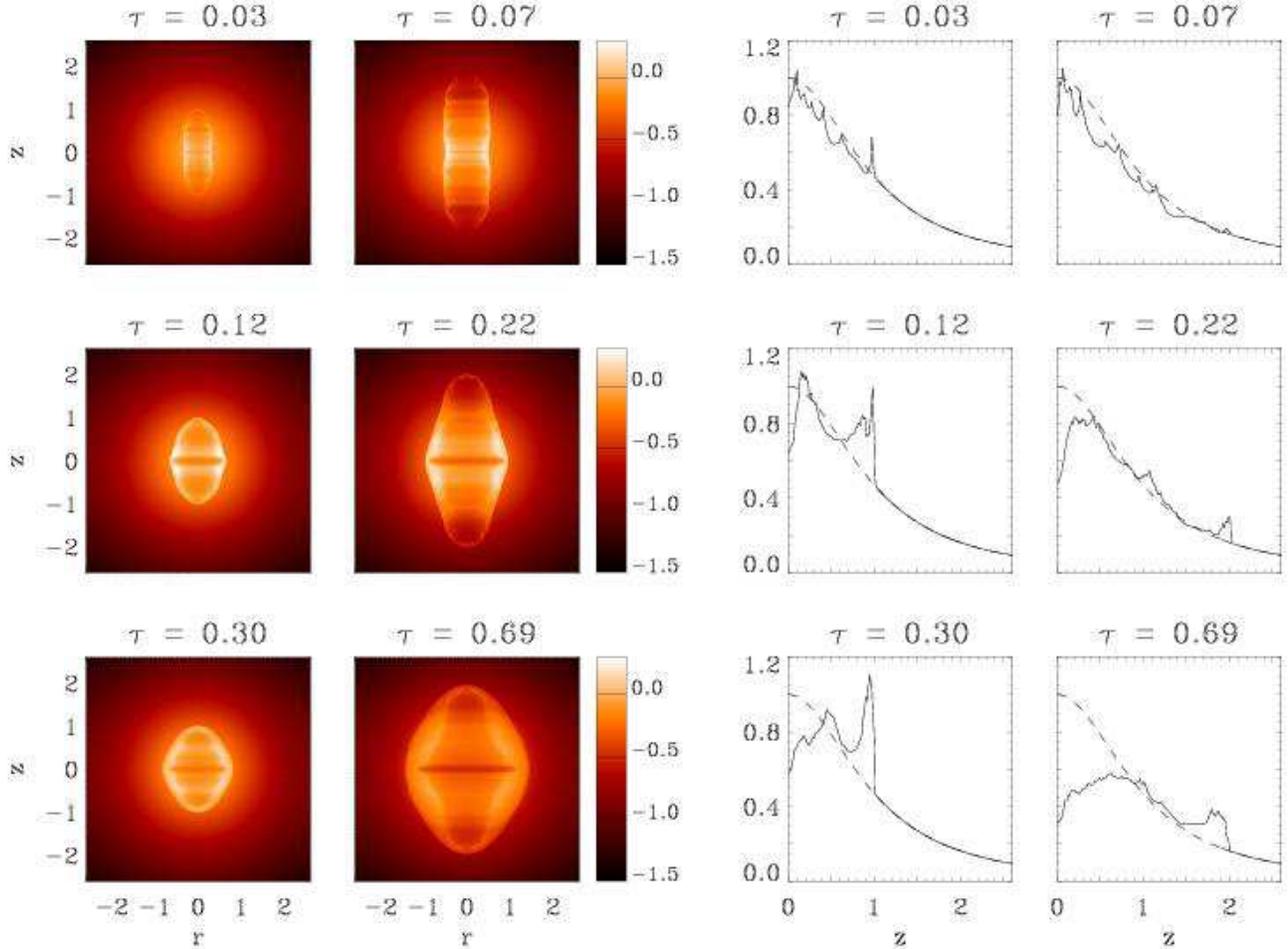}
  \caption{Simulated X-ray fluxes
   in the $0.1-4$ keV band for the $M=120$ cases. The rows
   refer from top to bottom to the $\nu=0.1,0.01,0.001$ cases respectively.
   In the first two columns on the left the X-ray images are shown in logarithmic scale at times
   corresponding to cocoon lengths of $l_{\rm c}\sim1$ and $2$ core radii.
   In the two columns on the right longitudinal cuts along to the jet axis corresponding
   to the images on the left
   are plotted with a solid line. The dashed line represents the emission from the undisturbed
   atmosphere. The fluxes are given in unity of the central flux of the
   unperturbed atmosphere}

 \label{fig1}
\end{figure*}

\begin{figure*}
\centering
  \includegraphics[width=17cm]{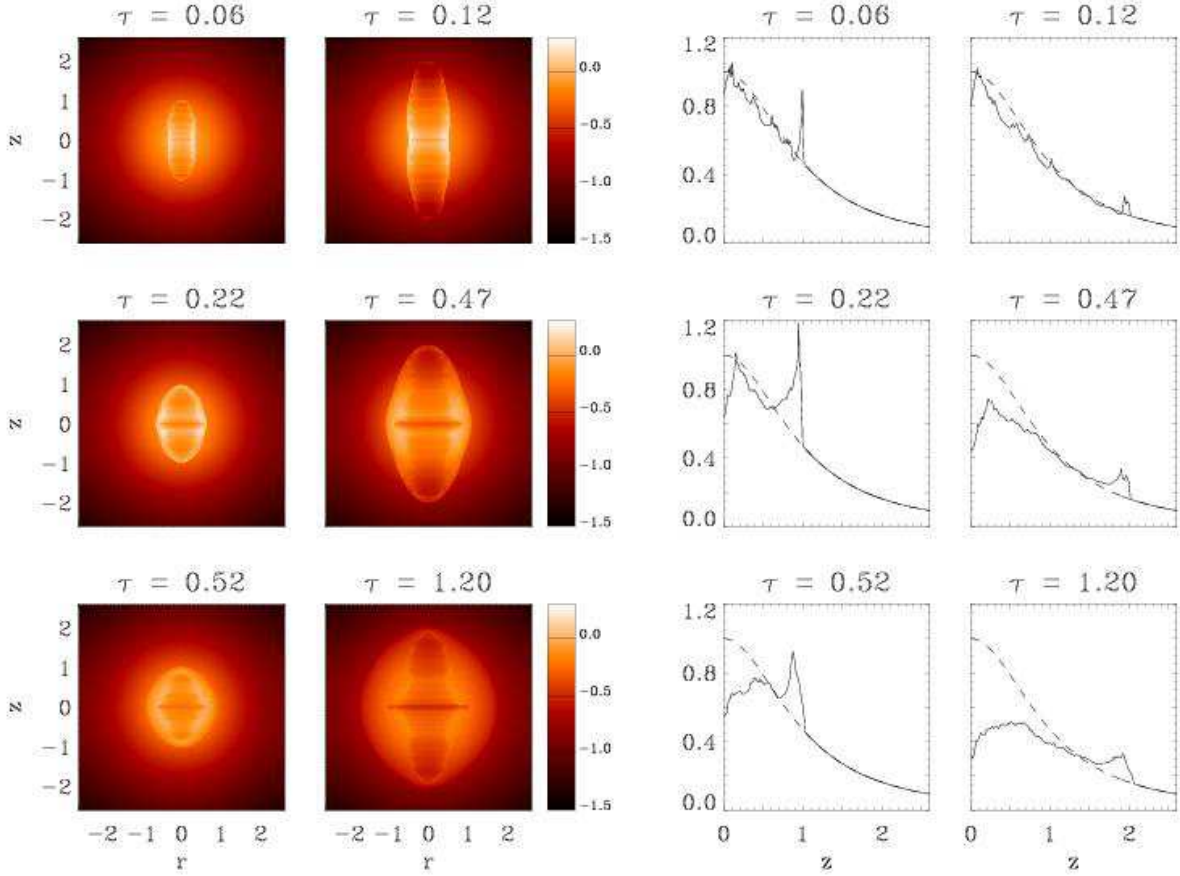}
  \caption{Simulated X-ray fluxes,
  in logarithmic scale, in the $0.1-4$ keV band for the $M=60$ cases. The rows
   refer from top to bottom to the $\nu=0.1,0.01,0.001$ cases respectively.
   The quantities represented are the same as in Fig. \ref{fig1}}
\label{fig2}
\end{figure*}

\begin{figure*}
\centering
  \includegraphics[width=17cm]{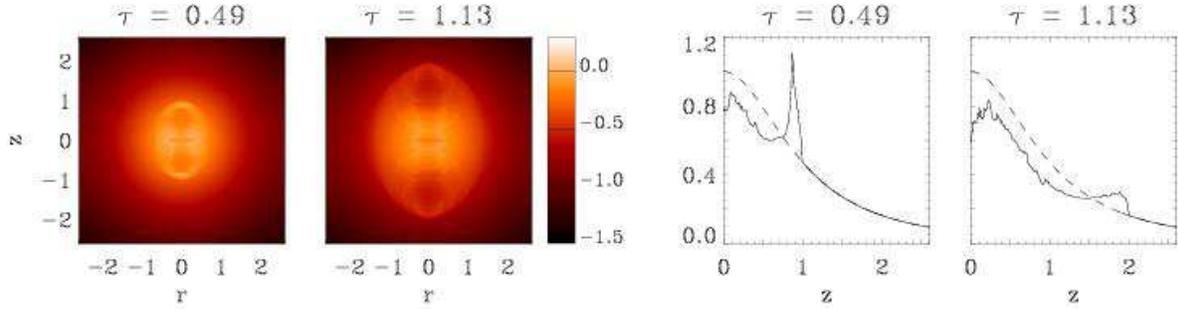}
  \caption{Simulated X-ray fluxes,
 in logarithmic scale,  in the $0.1-4$ keV band for the $M=10$, $\nu=0.1$ case.
   The quantities represented are the same as in Fig. \ref{fig1}-\ref{fig2}}
\label{fig3}
\end{figure*}

In Figs. (\ref{fig1} - \ref{fig3}) we show in the first
two columns simulated X-ray flux distribution
for the seven cases described in Table \ref{table:cases}.
The figures are symmetrical with respect to the $r$ and $z$ axes.
The emissivity per frequency unit is computed with a Raymond-Smith
thermal spectrum code (see HEASARC Web Page) from the electron density
and temperature distribution obtained in our calculation.
The emissivity $\epsilon=n^2\Lambda(T)$ is then
integrated in the $0.1 - 4 {\rm \;keV}$ band. The flux distribution
is calculated integrating the optically thin emissivity along the line of sight $\lambda$ 
that is assumed perpendicular to the jet axis:
\[
f=\int_{- r_{\rm domain}}^{+ r_{\rm domain}} \epsilon(n,T,\lambda) d \lambda \; .
\]
We consider an isothermal ambient medium with a temperature
$T = 2.3$ keV  and a central electron density $n_0 = 0.01\,{\rm cm}^{-3}$.
In Fig \ref{fig1} we show the results for the three cases
with $M = 120$, each row refer to a different value of the density
ratio and more precisely the top row is for $\nu = 0.1$, the middle
row is for $\nu = 0.01$ and the bottom row is for $\nu = 0.001$. The
first two panels in a row are for a different time and we have chosen the times in
order to have lengths respectively of $1$ and $2$ core
radii. Figure \ref{fig2} has the same structure, but is for the cases
with $M = 60$, while Fig. \ref{fig3} is for the single case with
$M = 10$ and $\nu = 0.1$.
A more quantitative view of these results can be obtained presenting
cuts of these figures along selected directions. In particular in
Figs. (\ref{fig1} - \ref{fig3}) in the third and fourth column we show
longitudinal cuts of the images along the jet axis. The figures correspond
to the images of the first two columns and refer
respectively to $M = 120$, $M = 60$ and $M = 10$. The dashed curves in all
the figures represent the emission from the undisturbed atmosphere.

These figures show a sequence of morphologies starting from cases in
which a cavity is not present going to cases in which the cavity is
the dominant feature of the image. In particular, we see that for $M
= 120$ the brightness depression is evident only for the lighter case,
while for the $M = 60$ cases starts to be present for $\nu = 0.01$
during its evolution and is very
evident during  the whole evolution for $\nu = 0.001$. For $M = 10$
the cavity is dominant already for $\nu = 0.1$.

The presence or absence of the brightness
depression, as discussed by Clarke, Harris \& Carilli (1997) depends on the
thickness of the shell. In fact, the line of sight that goes through
the cocoon region with very low emissivity crosses also the shell of enhanced
emissivity, and the observed flux is higher or lower than the undisturbed
profile depending on the interplay between the two effects. If the
shell is narrow, as a consequence of mass conservation, it will have
a high density,  the emissivity will be greatly enhanced and will overcome
the decrease in the cocoon giving an observed brightness higher than the
undisturbed profile. Following Clarke, Harris \& Carilli (1997), we can write the
ratio of the observed flux to the undisturbed one as
\beq
\frac {f'} {f} = \frac {1} {\delta (2 - \delta)^2}
\label{eq:fluxratio}
\quad,
\eeq
where we have neglected the dependence of emissivity on the temperature and
$\delta$ is the ratio between the shell width and the cocoon radius.
From Eq. (\ref{eq:fluxratio}), we see that the ratio increases as $\delta$
goes to 0 and becomes less than 1 for $\delta > 0.38$.

\begin{figure*}
\centering
  \includegraphics[width=17cm]{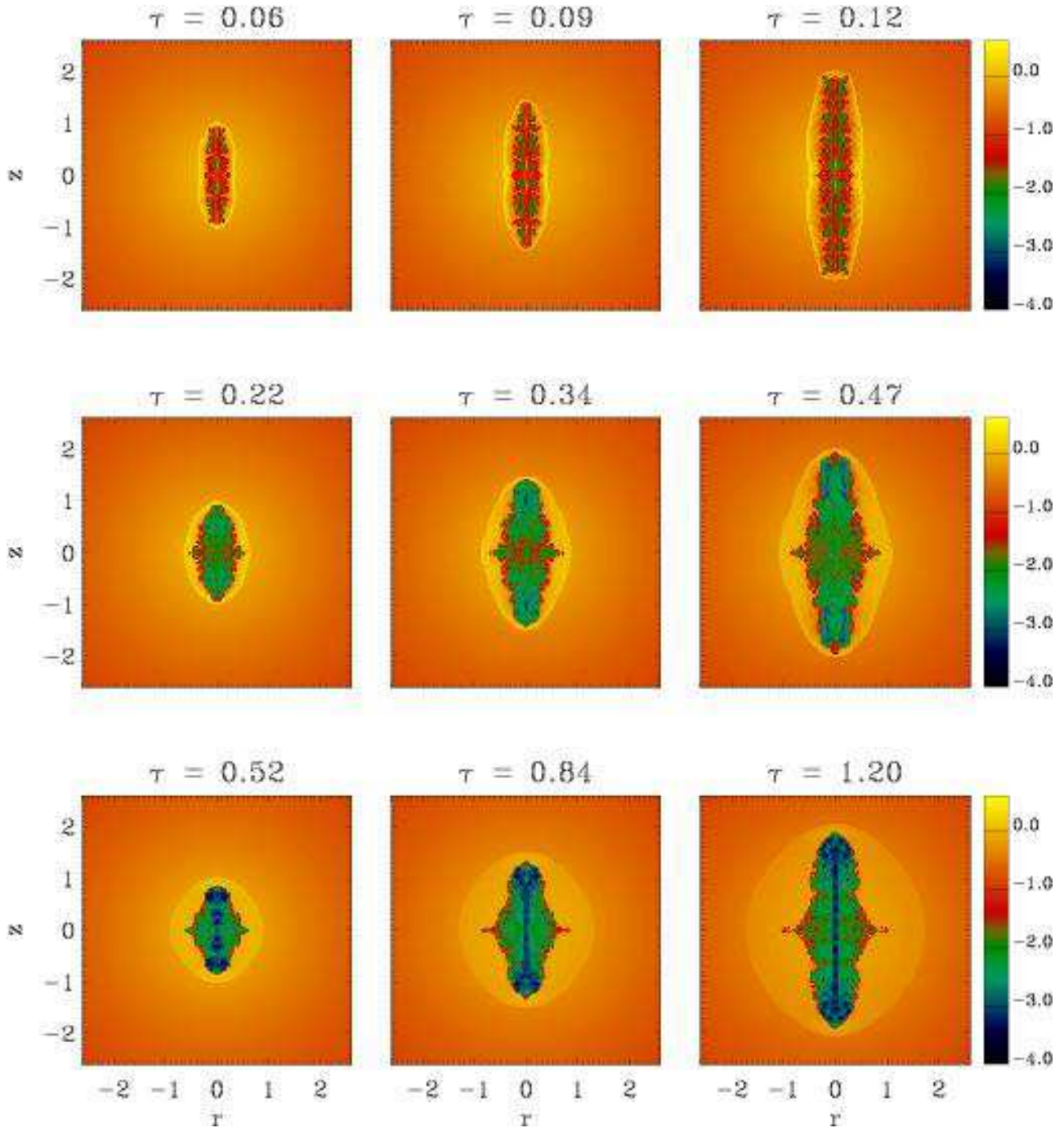}
  \caption{Density maps for the $M=60$ cases. The scale is logarithmic. The rows
   refer from top to bottom to the $\nu=0.1,0.01,0.001$ cases respectively.
   In the columns density maps are shown at times corresponding to a cocoon length
   $l_{\rm c} \sim 1,1.5$ and $2$ core radii. The densities are given in unity of the
   central density of the unperturbed atmosphere}

  \label{figdens}
\end{figure*}

In Fig. \ref{figdens} we show the
density distribution for the $M = 60$ cases: each row refer to a
different density ratio ($\nu =0.1, 0.01, 0.001$ from top to bottom)
and each column is for different times corresponding to cocoon lengths
$1, 1.5, 2$ respectively.
We see in fact that the width of the shell
increases going from the high to the low $\nu$ case. Moreover we can
see that the shell tends to widen and the bow shock becomes weaker
during the evolution of the cocoon, a part from the $\nu = 0.1$ case
in which the relative thickness of the shell remains constant.
 Note that the localized feature appearing along the $z=0$ axis
is due to the assumed reflection boundary conditions  and does not affect the
general morphological and dynamical behavior.

\begin{figure}[ht]
   \resizebox{\hsize}{!}{\includegraphics{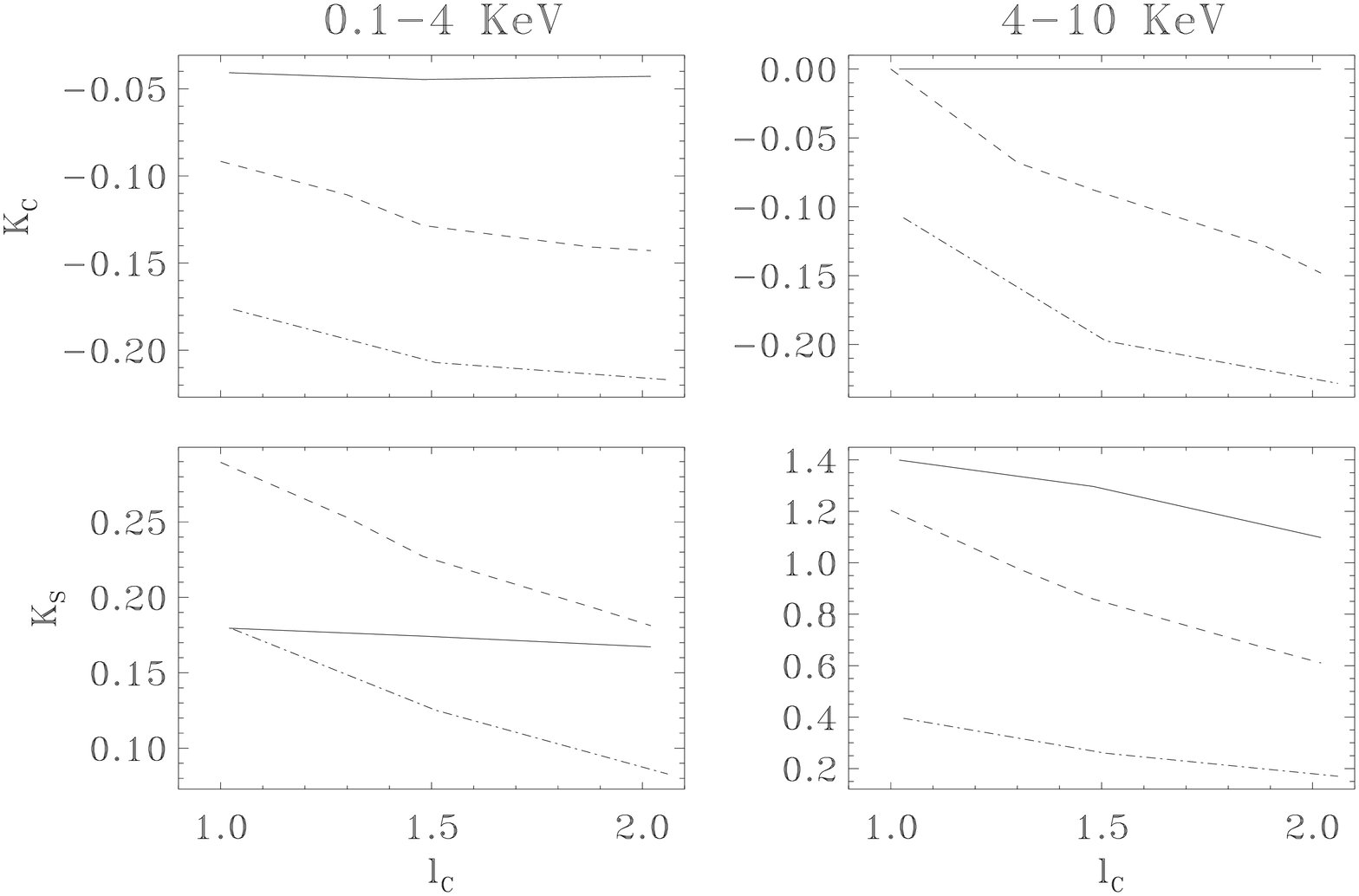}}
  \caption{({\it{Upper panels}}) Plot of the integrated flux deficit as a function
   of cocoon length in the $0.1-4$ keV band ({\it{left}}) and in the $4-10$ keV band
   ({\it right}) for the $M=60$ cases.
   This quantity is
   defined in Eq. (\ref{eq:cavity}).
   ({\it{Lower panels}}) Plot of the excess emission as a function of cocoon length in the
    $0.1-4$ keV band ({\it{left}}) and in the $4-10$ keV band
   ({\it{right}}) for the $M=60$ cases. This quantity is
   defined in Eq. (\ref{eq:shell}).
   In the four panels the
   three lines refer to the $\nu=0.1$ ({\it{solid}}), $\nu=0.01$ ({\it{dashed}})
   and $\nu=0.001$ ({\it{dash-dotted}}) cases}

  \label{fig:cavity60}
\end{figure}

%In order to give a more quantitative definition of the presence of a
%deficit in the X-ray emission  or at the opposite of an enhanced shell
In order to better quantify the deficit or the enhancement in the X-ray 
emission, we have computed integral measures $K_{\rm c}$ and $K_{\rm s}$ of these
quantities  defined as
\begin{equation}
  \label{eq:cavity}
   K_{\rm c} = \frac{1}{S_{\rm c}} \int_{S_{\rm c}}(f - f_{\rm k})dS
\end{equation}
\begin{equation}
  \label{eq:shell}
  K_{\rm s} = \frac{1}{S_{\rm s}} \int_{S_{\rm s}}(f - f_{\rm k})dS \; ,
\end{equation}
where the surface integrals are computed over the domains $S_{\rm c}$
and $S_{\rm s}$ defined as the areas over which the integrand $(f - f_{\rm k})$
is respectively lower (the cavity) and greater (the shell) than zero,
where $f$ is the flux at a given position and $f_{\rm k}$ is the flux at
the same position for an unperturbed King atmosphere.
The quantities $K_{\rm c}$ and $K_{\rm s}$ are calculated in two different energy bands
($0.1 - 4$ keV and $4 -10$ keV) and are plotted as a function of the length
of the cocoon $l_{\rm c}(t)$ in Fig. \ref{fig:cavity60}: the upper panels refer to $K_{\rm c}$,
while the lower panels refer to $K_{\rm s}$.  
In the figure we present the results for all the values of $\nu$
considered and for a single representative value of the Mach number $M
= 60$. In each figure the solid curve is for $\nu = 0.1$, the dashed
curve is for $\nu = 0.01$ and the dashed-dotted curve for $\nu = 0.001$.
Looking at the figure for $K_{\rm c} (0.1-4\,{\rm keV})$, we see that, as expected from
the results discussed above, in the case of $\nu = 0.1$ the
deficit is very low and does not increase with time,  instead, the deficit
increases at the decreasing of $\nu$ and shows an
increase with time. When we look at the excess emission in the $0.1-4$ keV band, measured by
$K_{\rm s}$, we see that the cases $\nu = 0.001$ and $\nu = 0.01$ behave as
expected, with the case $\nu = 0.01$ presenting an higher shell
emission with values decreasing in time (remember that the shell
widens and decreases its density during its evolution). The case $\nu
= 0.1$, instead, presents an unexpected behavior. In fact its value of
$K_{\rm s}$ is lower than that found for $\nu = 0.01$ and stays almost
constant with time. This however can be
understood by considering that the shock in the external medium heats
this medium changing its emission properties. The spectral range we
are considering for the flux calculation is $0.1-4$ keV and the
material in the shell may have been heated to a temperature for which
the maximum of the emission falls in a harder spectral range. These
effects will be however analyzed in more detail in the next Section
\ref{sec:temperature}.

\section{The shell temperature}
\label{sec:temperature}

\begin{figure}[ht]
    \resizebox{\hsize}{!}{\includegraphics{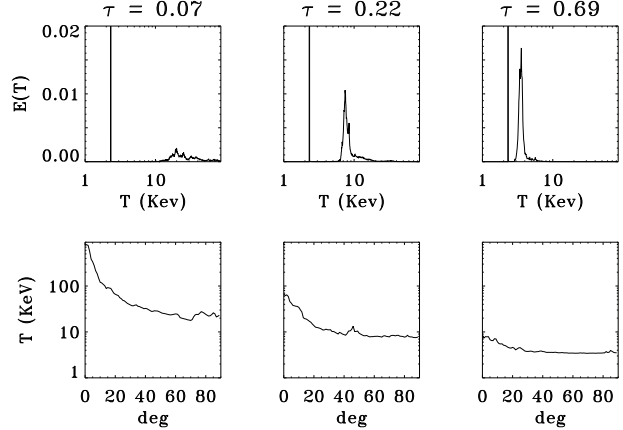}}
  \caption{Temperature plots
  for the $M=120$ cases ($\nu=0.1, 0.01, 0.001$ from left to right respectively)
  at a time corresponding to a cocoon length of $2$ core radii. ({\it{Upper panels}})
  Plot of the quantity $E(T)$ as defined in Eq. (\ref{eq:emiss}).
  It represents the energy emitted in the $0.1 - 10$ keV band by the gas at the temperature $T$.
  The vertical line marks the temperature 
  of the ambient isothermal medium while the peak correponds to the compressed shell emission.
  The emission is normalized to the total emission from the unperturbed atmosphere.
  ({\it Lower panels})
  Plot of the shell average temperature as a function of the angle
  with origin in $r=0, z=0$. The $0^\circ$ angle corresponds
  to the direction of the jet axis. The ambient temperature is taken $2.3$ keV}
  \label{fig:tempdistr120}
\end{figure}

\begin{figure}[ht]
    \resizebox{\hsize}{!}{\includegraphics{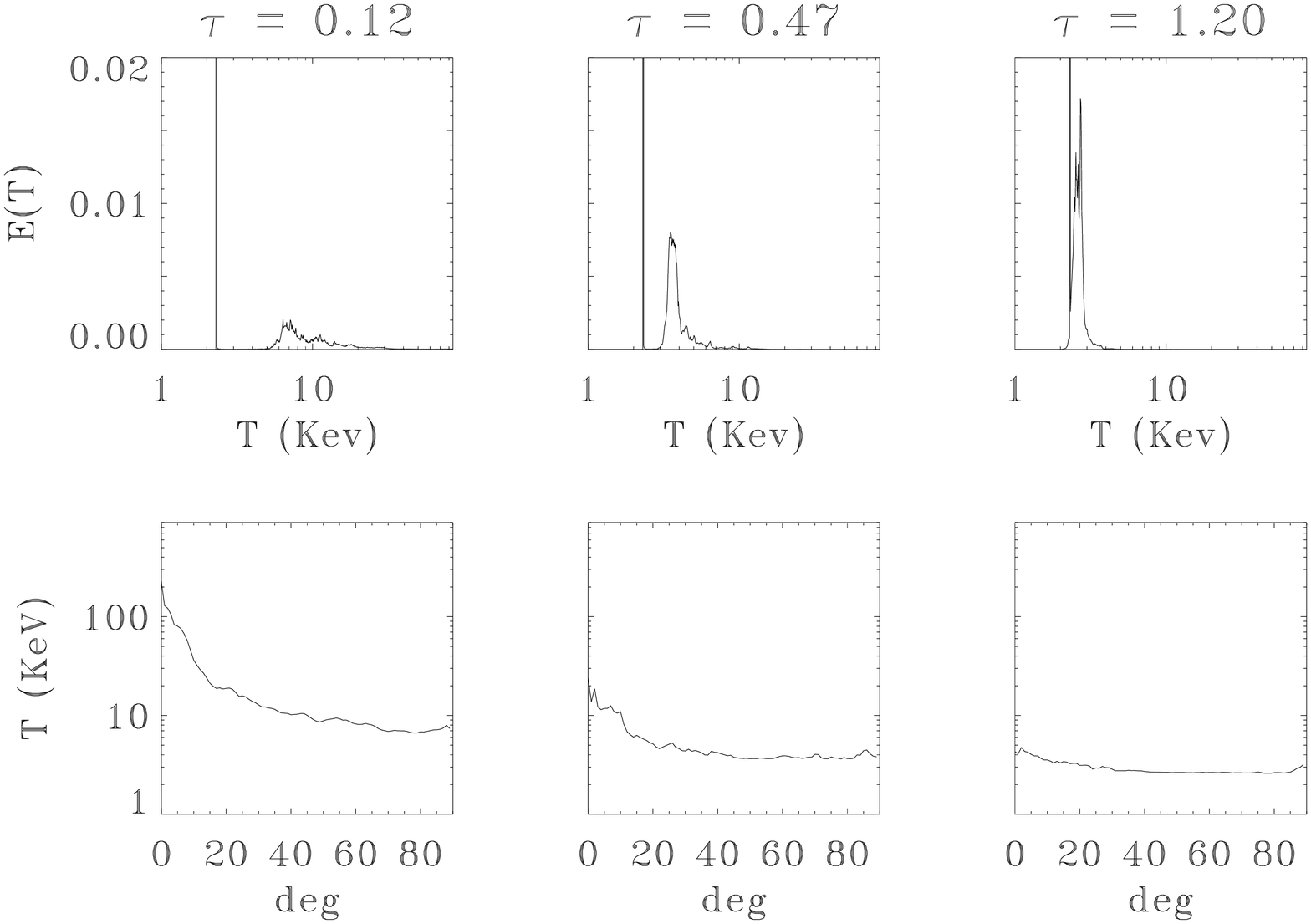}}
  \caption{Temperature plots
  for the $M=60$ cases ($\nu=0.1, 0.01, 0.001$ from left to right respectively)
  at a time corresponding to a cocoon length of $2$ core radii. The quantities plotted are the same
  as in Fig. \ref{fig:tempdistr120} }
  \label{fig:tempdistr60}
\end{figure}

\begin{figure}[ht]
%\sidecaption
    \includegraphics[width=3.5cm]{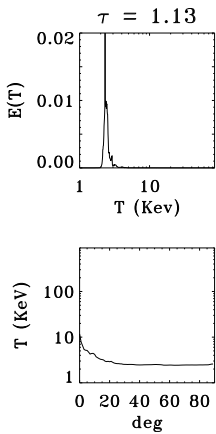}
%    \resizebox{\hsize}{!}{\includegraphics{figure8.eps}} 
  \caption{Temperature plots
  for the $M=10$, $\nu=0.1$ case at a time corresponding to a cocoon length of $2$ core radii.
  The quantities plotted are the same
  as in Fig. \ref{fig:tempdistr120} and in Fig. \ref{fig:tempdistr60}}
  \label{fig:tempdistr10}
\end{figure}

As we discussed, the shock driven in the external medium can heat it
and its effects can be more or less evident depending on its
strength. The heating of the shell can change the typical 
emission energies, an example of the consequences of this effect has
been shown in the bottom panel of the first column of
Fig. \ref{fig:cavity60} where we have seen that the case
with $\nu = 0.1$ seemed to present an anomalous behavior. We can now
compare the first column of Fig. \ref{fig:cavity60} with the second column where we
show the same quantities but in the range $4-10 $ keV. Looking at the
bottom panels we see that the case $\nu = 0.1$, which in the softer
band presented an excess emission below that of the case $\nu = 0.01$,
shows in the harder band an excess emission larger than the other cases.
The gas in the compressed shell becomes in fact hotter as we increase
the value of $\nu$ and the emission is then shifted towards higher
emission energies. To quantify in more detail this temperature change, we
have plotted in the first row of Figures \ref{fig:tempdistr120}-\ref{fig:tempdistr10}
the quantity
\begin{equation}
E(T) = \int_V \epsilon(n,T_1,{\bf x}) \delta(T_1({\bf x}) - T) d^3{\bf x} \; ,
\label{eq:emiss}
\end{equation}
where $T_1({\bf x})$ is the temperature at a given position,
$\delta(T_1-T)$ is the Dirac delta function and $\epsilon$
are the radiative losses per unit volume in the $0.1 - 10$ keV band as calculated
with the Raymond-Smith code.
This quantity measures the energy emitted by the gas
at the temperature $T$ in the $0.1 - 10$ keV band. In in the first row of Fig. \ref{fig:tempdistr120},
\ref{fig:tempdistr60} and \ref{fig:tempdistr10} we have plotted this quantity for
the Mach numbers 120, 60 and 10 respectively. In each figure columns are for 
different values of $\nu$ and the panels in each column are for
the time corresponding to a cocoon length of 2.
In each panel we see a vertical line marking the temperature
of the ambient isothermal medium and a peak correponding to the emission from
the shocked shell material.
We can see that the temperature of the shell increases as we
increase the Mach number and the density ratio $\nu$.
Taking into account the cases with the lowest density ratio $\nu=0.001$ we can see
that in the $M=120$ case the shell temperature is a factor two higher than the ambient one,
in the $M=60$ case it is slightly higher while in the $M=10$ case it is clear that
there is emission also from gas with a temperature lower than the ambient one.
This fact agrees with several observations (e.g. Perseus A) in which the coolest gas
is found in the shell of enhanced emission. The cooling in this simulation
is due to the adiabatic expansion of the shell after being compressed by a weak shock.
As time elapses, radiative losses may become important and contribute to the cooling
of the shell: for a central density of $0.04 \, {\rm cm}^{-3}$ and a temperature of $3 \, {\rm keV}$ (e.g. Perseus A, see Schmidt, Fabian \& Sanders 2002), the radiative cooling time is
$\sim 3\times 10^7$ years, i.e. close to the simulated time of the $M=10$, $\nu=0.1$ case.
The shell temperature can reach
temperature up to $800$ keV for the case $M = 120$ and $\nu = 0.1$.
The heating of the shell is not uniform but depends on the shock
strength and it is higher towards the jet head and becomes lower
further from the head. This is represented in the second row of
Figs. \ref{fig:tempdistr120}-\ref{fig:tempdistr10}, where we show the
average shell temperature as a function of angle for all the cases
considered.

From the figures we see that, for the cases with $\nu = 0.1$ and $M=60,120$, the
shell temperature  is everywhere larger than $8$ keV and reaches
temperatures up to $850$ keV at the jet head of the $M=120$ case. The lower density cases,
instead, present an increase in temperature mainly concentrated in the
forward part of the cocoon. This temperature distribution will have
consequences in the shell morphology as seen in different X-ray
bands. In fact, we expect to see emission from the forward part of the
cocoon at high energies, while the backward part is expected to be
more dominant at lower energies.

\section {The cocoon dynamics}
The numerical simulations presented in previous sections have shown
that the expanding cocoon drives a shock in the ambient medium, which
is then compressed and heated. The amount of compression and heating
is fundamental in determining the X-ray emission properties. The
results we have shown tell us that, as also discussed by Reynolds, Heinz \& Begelman 
(2001), the
strength of the shock, driven by the cocoon, weakens as the cocoon
expands and the stage at which the transition between a strong and a
weak shock occurs depends on the jet properties, i.e. its Mach number
and its density ratio. In this section we will try to determine in a
more quantitative way this dependence of the transition on the jet
physical properties. We will do that through a comparison between
analytical models for the cocoon expansion in an homogeneous medium
and the results of our numerical simulations. In the next subsection
we will then examine the cocoon dynamics in a uniform medium, in
subsection \ref{sec:stratified} we will examine the dynamics in a
stratified medium and in in subsection \ref{sec:transition} we will
try to determine the dependence of the transition discussed above on
the jet properties.

\subsection{Expansion in a uniform medium}

The first attempt to build an analytical model for the cocoon dynamics
is due to Begelman \& Cioffi (1989) and Cioffi \& Blondin (1992). They
considered only the case of strongly overpressured cocoons, for which
they consider only strong shocks driven in the external medium and
they essentially neglect the external pressure.  For our purposes, we
have to extend the description to a more general situation, in which
the external shock can be of any strength and the external pressure is
taken into account. We describe in detail the model in the Appendix,
here we give only the resulting behavior of the extended cocoon
radius and cocoon pressure versus time: the cocoon radius  is
given by
\begin{equation}
r_{\rm e}^2 \, = \, c_{se}^2 t^2 + \sqrt{\gamma^2-1} r_{\rm j} \nu^{1/4} M c_{se} t
\label{eq:trcocoon}
\end{equation}
and the cocoon pressure is
\beq
\frac{P_{\rm c}}{P_{\rm ext}} \, = \, 1+\frac{\gamma(\gamma-1)}{2}
\left(\frac{1}{\nu^{1/2} M^2}\frac{c_{se}^2 t^2}{r_{\rm j}^2}
+ \frac{\sqrt{\gamma^2-1}}{\nu^{1/4} M}\frac{c_{se} t}{r_{\rm j}}\right)^{-1} \, .
\label{eq:tpcocoon}
\eeq

Equations (\ref{eq:trcocoon}, \ref{eq:tpcocoon}) describe two
different phases in the evolution: initially the cocoon is strongly
overpressured and the solutions behave in the same way as described by
the Begelman \& Cioffi (1989) model, with the radius proportional to
$t^{1/2}$ and the pressure proportional to $t^{-1}$.
As the pressure decreases, the
contribution of the external pressure becomes more
important and the external shock becomes weak.
In this second phase the pressure tends to become constant  and the
extended  cocoon expands essentially at
the sound speed and increases as $\propto t$. 
We can use Eq. (\ref{eq:rci}) to determine the radius of the
cocoon proper: in the first phase the internal radius increase $\propto t^{1/2}$
like the external one not allowing the shell to expand, while
in the second phase it tends to a
constant value. The compression of the external medium can be related
to the relative shell thickness and we can
observe that, while it stays constant in the first phase, as already
discussed above, it tends to increase in the second phase.
The transition between these two phases happens when the two terms on the 
right hand side of Eq. (\ref{eq:trcocoon}) become comparable;
then, assuming a cocoon length $l_{\rm c}=\nu^{1/2}v_{\rm j}t$, we can determine
a scaling law for the length of transition $l_{\rm c}^*$ between the two regimes
\begin{equation}
  \label{eq:transl}
  l_{\rm c}^* \propto M^2\nu^{3/4} \; .
\end{equation}
Therefore, for a given Mach number and for a given
length of the cocoon, we expect the jets with lower density ratios to
be less overpressured and therefore to form wider and less dense
shells. 

In a similar way we can also find a solution assuming a spherical symmetry
for the cocoon, that is  solving the system of equations (\ref{eq:pressure2} 
- \ref{eq:vcocoon}) taking $V_{\rm c} = 4\pi /3 r_{\rm e}^3$. 
Taking into account separately the
strongly overpressured ($P_{\rm c} \gg P^*$) and the weakly overpressured
($P_{\rm c} \simeq P^*$) phases we can find two distinct regimes of expansion:
a supersonic one during which the external radius behaves like
\beq
r_{\rm e}=\left[ \frac{25(\gamma^2-1)}{9\pi}\right]^{1/5}
\left(\frac{L_{\rm j}}{\rho_{\rm ext}}\right)^{1/5}t^{3/5}
\label{eq:resph}
\eeq 
and the pressure decreases as
\beq
P_{\rm c}=\frac{3(\gamma-1)}{4\pi}\left[\frac{9\pi}{25(\gamma^2-1)}\right]^{3/5}
\left(\rho_{\rm ext}^3 L_{\rm j}^{2}\right)^{1/5}t^{-4/5}
\label{eq:prsph}
\eeq
and a weakly overpressured phase during which the cocoon pressure tends to a
constant value and the radius $r_{\rm e}$ expands $\propto t$ at the sound speed.
Assuming spherical symmetry also for the cocoon proper ($V_{\rm ci}=4\pi 3r_{\rm i}^3$)
we can derive the behavior of the cocoon radius $r_{\rm i}$ solving Eq. (\ref{eq:vci}):
$r_{\rm i}$ expands as $t^{3/5}$ during
the strongly overpressured phase, with the same rate of expansion of $r_{\rm e}$,
while it behaves as $t^{1/3}$ during the weakly overpressured one. As in the cylindrical
geometry model the shell is allowed to widen only in the weakly overpressured phase. 
We can then determine a scaling law for the radius of transition between the two regimes
imposing that the expansion speed determined by Eq. (\ref{eq:resph}) approaches the
sound speed $c_{\rm se}$  
\beq
r_{\rm e}^* \propto \left(\frac{L_{\rm j}}{c_{\rm se}^3\rho_{\rm ext}}\right)^{1/2} \; .
\label{eq:trasph}  
\eeq 
In the spherical symmetry model the cocoon properties scale with the jet power since it
is the only jet parameter that enters the system of equation solved in the Appendix. 
It is worthwhile noticing that our solution for the spherical strongly overpressured cocoon 
(Eq. \ref{eq:resph} and \ref{eq:prsph}) is the same, apart from numerical constants, of
Heinz, Reynolds \& Begelman (1998) who solved a system of equations similar to ours but limited their 
solution to the strongly overpressured regime.

Looking at the geometry and at the detailed structure of the shell and the cocoon, these analytical models 
are too simplified  to describe in detail the radiative properties of the cocoon but still they show clearly how
the relative thickness of the shell can increase only in a weakly overpressured regime.

\subsection{Expansion in a stratified medium}
\label{sec:stratified}

The cocoon expansion in a stratified medium is, of course, much more
complicated, since the external pressure is not constant, and the
development of an analytical model becomes more problematic.  However
the above discussion can provide a framework for understanding the
results of numerical simulation. 

\begin{figure}[ht]
   \resizebox{\hsize}{!}{\includegraphics{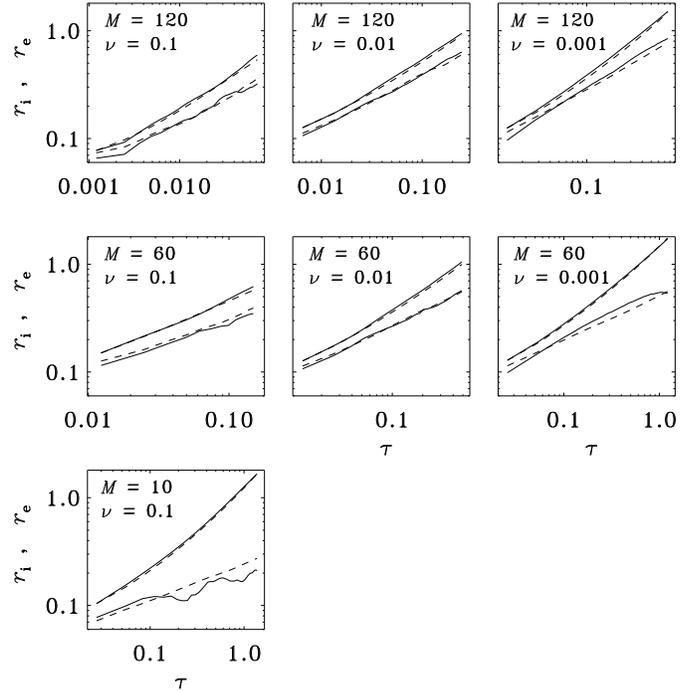}}  
   \caption{Plot of the internal and external radii vs time. The rows refer to the 
   different Mach cases while columns to the different $\nu$ values. The solid lines
   show the extended and internal cocoon radii as determined in the simulations
   taking an average value between $1/8$ and $1/4$ of the total length of the cocoon.
   The dashed lines show the same quantities as determined by our analytical model assuming
   that the expansion of the cocoon is driven by the average pressure
   of the cocoon}
   \label{fig:rcocoon}
 \end{figure}

We recall that the presence of a brightness depression depends 
substantially on the relative thickness of the shell. We then must first study
the evolution of the extended cocoon radius and the cocoon proper one.
We first analyze the behavior of the extended cocoon radius $r_{\rm e}$ by
comparing the results directly obtained by the numerical simulations
with an estimate obtained, using the average cocoon pressure as the
driver for the expansion. More precisely, we obtain this estimate by
integrating numerically Eq. (\ref{eq:vcocoon}), where for $P_{\rm c}(t)$
we use the average value obtained at every time step from the simulations. For defining
the extended cocoon radius in the simulations, we concentrate on the
base portion of the cocoon and we take an average value of the bow
shock radius between 1/8 and 1/4 of the total length of the cocoon.
The results of the comparison are represented in  Fig. \ref{fig:rcocoon},
where we plot $r_{\rm e}$ and $r_{\rm i}$ as a function of time for all the cases we have
considered. In all the panels the solid curves correspond to the
results obtained in the simulations, while the dashed curves correspond
to the estimates obtained through the average pressure. The figures show 
that this estimate reproduces very well the actual behavior of the extended cocoon
radius: 
this result tells us that the average cocoon pressure is a good estimate
of the local pressure driving the cocoon expansion.
In a similar way we can proceed for the cocoon
radius $r_{\rm i}$, for which
we can use eq. (\ref{eq:vci}) where with $\eta L_{\rm j}$ we intend
the fraction of jet power that is thermalized and goes into internal energy of the cocoon.
This fraction can be estimated from eq. (\ref{eq:pressure2}). 
Eq. (\ref{eq:rci}) can be integrated giving
\begin{equation}
  \label{eq:vcocoon}
  V_{\rm c} = \frac{1}{{P_{\rm c}(t)}^{1/\gamma}}\left\{
P_{\rm c}(0)^{1/\gamma}V_{\rm c}(0)+
\frac{\gamma-1}{\gamma}\int_{t_0}^t \eta L_{\rm j}
P_{\rm c}^{(1-\gamma)/\gamma}dt\right\}  
\end{equation}
where again for $P_{\rm c}$ we can use the average cocoon pressure obtained
above. This equation gives us an estimated behavior of the cocoon
volume and from this we can estimate the behavior of the cocoon radius
as $r_{\rm i} = \sqrt{V_{\rm c}/K l_{\rm c}}$ where $K$ is a form factor that depends on
the shape of the cocoon. One problem in determining $r_{\rm i}$ is that the
shape of the cocoon is strongly variable since the interface between
cocoon and external shocked material is subject to Kelvin-Helmholtz
instabilities. Moreover, the shape of the cocoon may depend on the jet
physical parameters. However, we can see from Fig. \ref{fig:rcocoon}
(where, as before with $r_{\rm e}$,  we compare this estimate with the actual
results obtained from the simulations)
that using an appropriated average values for $K$, we can capture
quite well the average behavior. The form factor should be $K=\pi$ for
a cylinder and $K=2\pi /3$ for a hemisphere. The value that can be
estimated from the simulations and that works quite well for every case
is $K\sim 2$. Again, this good agreement between
the estimates obtained through the usage of the average cocoon
pressure and the actual results tells us that we can use the average
pressure behavior for understanding the cocoon dynamics

Looking in more detail at the behavior of the extended cocoon radius,
we observe that, as expected from the discussion of the uniform case,
its expansion velocity decreases faster initially, when the cocoon is
strongly overpressured, and then decreases more slowly, when the shock
becomes weaker. Fitting a power law $\propto \tau^{\alpha}$ for the
quantities $l_{\rm c}$, $r_{\rm e}$, $r_{\rm i}$ in the initial and
terminal part of the evolution for the cases with $M = 60$, we get the
exponents $\alpha$ reported in Table \ref{tab:exponents}. We see
that, at later times, the exponents of $r_{\rm e}$ become systematically larger than those
obtained at the initial times.  Moreover we see that the differences
between the two exponents are larger for the lower density cases and
for $\nu = 0.1$ are minimal. In this last case, in fact the cocoon
stays strongly overpressured during the whole evolution.

\begin{table}[ht]
  \begin{center}
    \caption{The Table shows the exponent $\alpha$ of the fitted power laws $\tau^\alpha$
             for the quantities $l_{\rm c}$ (cocoon length), $r_{\rm e}$ (extended cocoon radius) and
             $r_{\rm i}$ (internal cocoon radius). The left column refers to the initial phase
             of the evolution of the cocoon ($l_{\rm c} <1.0$) while the second one to the
             advanced one ($l_{\rm c} >1.5$)}

\begin{tabular}{ccc}
  \hline \hline
  $\nu$ = 0.1 & $l_{\rm c}$ $<$ 1.0 & $l_{\rm c}$ $>$ 1.5 \\ \hline
  $l_{\rm c}$ & 0.85 & 0.85 \\
  $r_{\rm e}$ & 0.54 & 0.67 \\
  $r_{\rm i}$ & 0.43 & 0.56 \\ \hline

  \hline \hline
  $\nu$ = 0.01 & $l_{\rm c}$ $<$ 1.0 & $l_{\rm c}$ $>$ 1.5 \\ \hline
  $l_{\rm c}$ & 0.75 & 0.95 \\
  $r_{\rm e}$ & 0.56 & 0.71 \\
  $r_{\rm i}$ & 0.45 & 0.53 \\ \hline

  \hline \hline
  $\nu$ = 0.001 & $l_{\rm c}$ $<$ 1.0 & $l_{\rm c}$ $>$ 1.5 \\  \hline
  $l_{\rm c}$ & 0.77 & 0.89 \\
  $r_{\rm e}$ & 0.69 & 0.81 \\
  $r_{\rm i}$ & 0.46 & 0.07 \\ \hline
\end{tabular} 
 
    \label{tab:exponents}
  \end{center}
\end{table}

A similar analysis can be done for the cocoon radius $r_{\rm i}$. We see that
in the high and intermediate density cases the exponent for $r_{\rm i}$ also
increases but less than the exponent for $r_{\rm e}$. In the low density
case, instead, $r_{\rm i}$ becomes almost constant in time. This behavior
tells us that, when the cocoon is not any more strongly overpressured,
the relative shell thickness start to increase as it happened in the
homogeneous case.  The behavior of the relative thickness of the shell
can be in fact derived from Table (\ref{tab:exponents}), since it is
related to the ratio $r_{\rm i}/r_{\rm e}$. We then see that it has in general a
slower increase at the beginning of the evolution, and accelerates in
the following phases. As discussed above, the effects of not being
strongly overpressured are more evident by decreasing the value of
$\nu$ and, in fact, the increase of the relative cocoon width is larger
for smaller $\nu$. If we compare the cocoon widths for equal distances
of jet propagation we have a further effect that amplify the
consequences of the behavior discussed above, namely we have to take
into account that the advance velocity of the head of the jet
decreases when we decrease the value of $\nu$. 

In this subsection we have seen that the transition from a strongly
overpressured cocoon to a cocoon which is essentially in pressure
equilibrium with the surroundings leads to different properties of the
shell of the compressed external material. In the next subsection we
will try to see whether it is possible to determine in a more
quantitative way how this transition depends on the jet physical parameters.

\subsection{The transition from strong to weak shocks}
\label{sec:transition}

From the above discussion we have seen that the average cocoon
pressure can be used for interpreting the cocoon dynamics and therefore
determining the transition between the strong and weak shock regimes.
In the  case of uniform external medium, the cocoon
pressure, after a decrease proportional to $t^{-1}$, tends to a
constant. In the present case, with a decreasing external
pressure, we expect that the cocoon pressure does not tend to a
constant but that it will tend to follow the behavior of $P^*$,
defined in eq. (\ref{eq:pstar}).

 \begin{figure}[ht]
   \resizebox{\hsize}{!}{\includegraphics{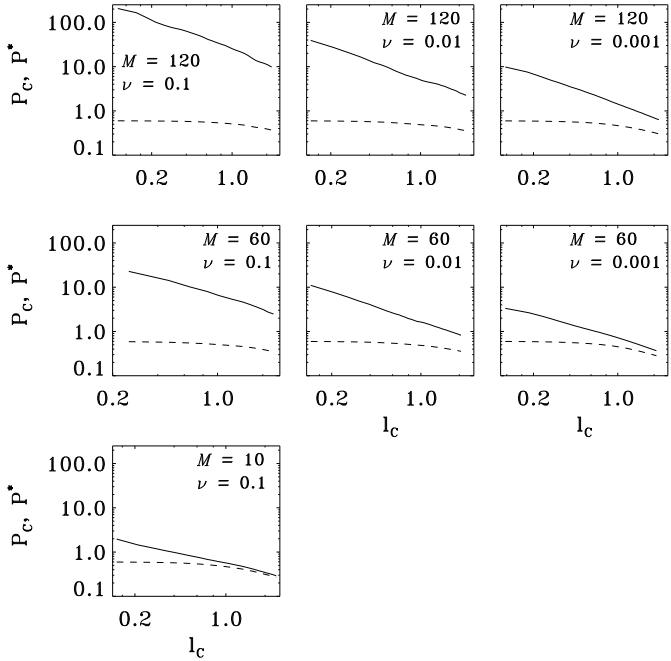}}
   \caption{Plot of the pressure vs cocoon length. The rows refer to the
   different Mach numbers while columns to the different $\nu$ values. The solid line shows the
   evolution of the average cocoon pressure with time in the different cases. The
   dashed one represents the external pressure averaged on the cocoon volume as defined in
   Eq. \ref{eq:pstar} }
   \label{fig:pressure}
 \end{figure}

 In Fig. \ref{fig:pressure} we have plotted the behavior of
 $P_{\rm c}$ (solid curve) and $P^*$ (dashed curve) against the cocoon length, where $P_{\rm c}$ is the
 average cocoon pressure and $P^*$ is the external pressure averaged
 over the cocoon volume. The ratio between the two quantities gives a
 measure of the average strength of the shock driven in the external
 medium.  
Looking at the behavior of the cocoon evolution in the different cases  we
have an always strongly overpressured expansion for the cases $M=120$ and $ \nu=0.1, \  0.01$
and for $M=60$ and $\nu=0.1$, in these three cases the transition length
would be larger than the actual longitudinal size of our domain; two cases showing
a transition from strongly overpressured to weakly overpressured cocoons for $M=120$, $\nu=0.001$ and
$M=60$, $\nu=0.01$; finally we have two cases where the cocoon is
always weakly overpressured,
i.e. for $M=60$, $\nu=0.001$ and $M=10$, $\nu=0.1$. This is in agreement with
the discussion on the X-ray flux distributions following  Figs. (\ref{fig1} - \ref{fig3}).

\begin{figure}[t]
   \resizebox{\hsize}{!}{\includegraphics{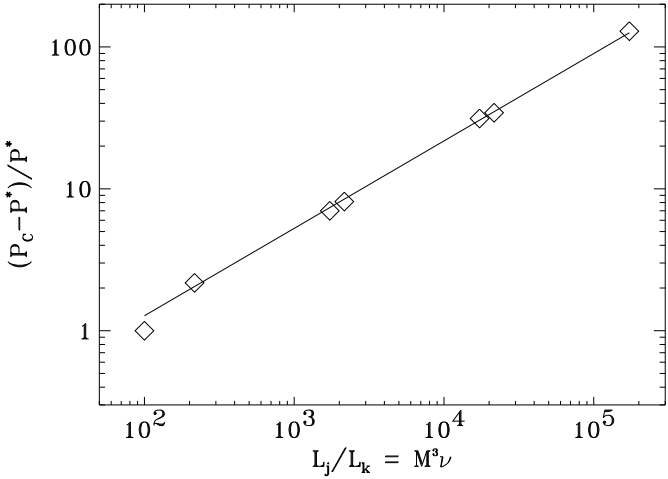}}
   \caption{Plot of the values of $(P_{\rm c} - P^*)/P^*$ at a cocoon length $l_{\rm c} = 0.3$
            (see Table \ref{table:pressure}) against jet kinetic luminosity
             $L_{\rm j}/L_{\rm k}=M^3\nu$. The solid line is a $(P_{\rm c} - P^*)/P^* \propto (L_{\rm j}/L_{\rm k})^{0.62}$,
             that is also the best fit, excluding the value of the case $M=10$, $\nu=0.1$ that is
             in a weakly overpressured regime already}
   \label{fig:scallum}
 \end{figure}

In order to determine the length of the cocoon at which the transition occurs, we look for
%In order to determine the expansion conditions for the different cases, we look for
a scaling law for the quantity $(P_{\rm c} - P^*)/P^*$ as a function of the parameters
$l_{\rm c}$, $M$ and $\nu$ during the initial strongly overpressured phase.
Taking into account only the cases that are stronlgy overpressured at the beginning of the evolution,  
we notice that the  initial decrease of the average cocoon pressure follows a similar
 behavior for all the cases that we have considered:
 with a power law fit $(P_{\rm c} - P^*)/P^* \propto l_{\rm c}^{- \alpha}$ to the
 initial evolution of the different cases we find a mean value $\alpha=0.9$ with
 an uncertainness of $10\%$.
 In Table \ref{table:pressure} we show the
 values of the quantity $(P_{\rm c} - P^*)/P^*$ at the beginning of the evolution,
 when the cocoon length is one third the core radius and the cocoon is
 typically strongly overpressured (notice that the cases
 $M=10$, $\nu=0.1$ and $M=60$, $\nu=0.001$ make an exception). These values can be
 represented, with good approximation by the scaling $(P_{\rm c} - P^*)/P^* \propto
 M^{1.85} \nu^{0.62}$. The general scaling law thus becomes:
 \beq
 \frac { P_{\rm c} - P^* } {  P^*}   \propto l_{\rm c}^ { -0.9}M^{1.85}\nu^{0.62}   \,.
 \label{eq:scal}
 \eeq
 The scaling relation (\ref{eq:scal}) shows that the initial pressure
 of the cocoon is not proportional to an arbitrary combination of the jet parameters
 $M$ and $\nu$ but it scales with the normalized kinetic jet power
 $L_{\rm j}/L_{\rm k} = M^3 \nu$ (see Eq. (\ref{eq:L_j})) as $(P_{\rm c} - P^*)/P^* \propto l_{\rm c}^ { -0.9}(L_{\rm j}/L_{\rm k})^{0.62}$.
 This result is confirmed by the plot of the values of $(P_{\rm c} - P^*)/P^*$ reported in
 Table \ref{table:pressure} against the jet kinetic luminosity $L_{\rm j}/L_{\rm k}$,
 after setting $l_{\rm c}=0.3$. The plot is shown in Fig. \ref{fig:scallum}.

%\begin{table}[ht]
%\begin{center}
%\begin{tabular}{l|c|c|c}
%\label{table:pressuredecay}
%M \textbackslash \ $\nu$  & 0.1  & 0.01 & 0.001 \\ \hline
%120      & 0.85 & 1.01 & 0.98   \\
%60       & 0.83 & 1.04 & 1.01  \\
%10       & 1.25 &      &
%\end{tabular}
%\caption{Exponents $\alpha$ of the fitted power law $(P_{\rm c} - P^*)/P^* \propto 
%l_c^{-\alpha}$ for the different simulations. The first phase of evolution
%is considered ($l_c<1$).}
%\end{center}
%\end{table}

\begin{table}[ht] 
\begin{center}
\caption{Values of the quantity $(P_{\rm c} - P^*)/P^*$ for the different simulations
evaluated at a cocoon length $l_{\rm c} =0.3$}
\begin{tabular}{c|ccc}
\hline \hline
$M$ \textbackslash \ $\nu$  & 0.1  & 0.01 & 0.001 \\ \hline
120      & 192.03 & 31.30 & 6.99   \\
60       & 34.35 & 8.13  & 2.17  \\
10       & 0.99 &      & \\ \hline
\end{tabular}
\label{table:pressure} 
\end{center}
\end{table}

Of course the transition from a strongly
to a weakly overpressured  regime  is not sudden, but gradually happens when  $(P_{\rm c} - P^*)/P^*$
becomes of the order of unity. Setting a constant value for $(P_{\rm c} - P^*)/P^*$ of this order
in Eq. (\ref{eq:scal})
we derive the scaling law for  transition length $l_{\rm c}^*$   (as we have done
for the uniform case, see Eq. (\ref{eq:transl})).

\begin{equation}
  \label{eq:ltransition2}
  l_{\rm c}^* \propto M^{2.05} \nu^{0.69}
\end{equation}
or in terms of the jet kinetic power
\begin{equation}
  \label{eq:ltransition3}
  l_{\rm c}^* \propto \left(\frac{L_{\rm j}}{L_{\rm k}}\right)^{0.69} \; .
\end{equation} 

Equation (\ref{eq:ltransition2}) represents a family of curves in the plane ($M,
\nu$), and, for a given cocoon length one of the curves of this family
will mark the separation between cocoons that drive strong shocks in
the ambient medium (strongly overpressured regime) and cocoons that drive only weak shocks
(weakly overpressured regime).
In the next
section we will discuss the astrophysical relevance of these results.

\section{Discussion}

\begin{figure}
  \resizebox{\hsize}{!}{\includegraphics{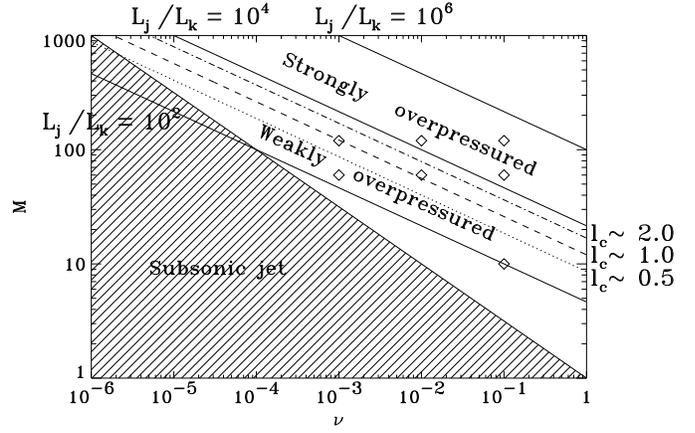}}
  \caption{Representation of the different regimes of the cocoon expansion in the
   $M - \nu$ plane. Inside the dashed area the jet becomes subsonic with respect to
   its internal sound speed. The solid lines correspond to a constant kinetic jet power 
   $L_{\rm j}/L_{\rm k}=M^3\nu$ given in unit of the power $L_{\rm k}$ defined in Eq. (\ref{eq:Lk}). The
   fragmented curves separate strongly and weakly overpressured regimes for different
   lengths of the cocoon given in core radii units. It can be seen that these lines correspond to
   lines of constant kinetic jet power. The diamonds correspond to the simulated cases
     }\label{mnuplane}
\end{figure}

In the previous sections we have seen that strongly
and weakly overpressured cocoons present different X-ray
morphologies. The former ones will show no deficit of emission
accompanied by a strong
emission from a shell marking the shock, driven by the cocoon
expansion. The material in the shell will be much hotter than the
ambient medium, the shell emission will then be shifted to higher
frequencies and therefore more visible at higher
X-ray energies, especially near the jet head. Weakly
overpressured cocoons will be instead characterized by the presence of a
deficit of emission in the cocoon, while the emission from the shell will be much
less visible than in the previous case; in addition, the material in the shell will be
essentially at the same temperature as the ambient medium and there
will be no change in the emission spectrum. These two regimes depend
on the jet physical parameters and on the age of the cocoon.
High Mach
number jets with high densities will be strongly overpressured for a
longer fraction of their life, decreasing the density and the Mach
number the transition to weakly overpressured cocoon will occur at an
earlier stage. We have found that the scaling  of the transition
length from one regime to the other is given to a good approximation by
Eq. (\ref{eq:ltransition2}). Fixing a length of the cocoon, from this
equation we can then obtain a relation between Mach number and density
ratio such as $M = C(l_{\rm c}) \nu^{-1/3}$.  This scaling law has been
represented in Fig. \ref{mnuplane} with the dotted, dashed and 
dash-dotted lines for cocoon lengths equal to 0.5, 1 and 2 core
radii respectively. The proportionality constant $C(l_{\rm c})$ has been set
to fulfill the transition conditions in the cases where the
transition between the two regimes is observed, i.e. $M=60, \ \nu=0.01$
and $M=120, \ \nu=0.001$. In the same figure we can
also notice the presence of another line, that individuates a region in
which the jet becomes subsonic with respect to its internal sound
speed. 

Since Mach number and density ratio are physically very meaningful
parameters, but cannot be determined for actual jets, it is favorable
to express the scaling relation in terms of the jet kinetic power.
In order to do so, we make use of Eq. (\ref{eq:ltransition3})
where $L_{\rm k}$ is a quantity that depends on the properties of the
environment. From this relation we see that jets with the same
kinetic power have the same transition  length
between the two phases. The curves in Fig. \ref{mnuplane}
separating the strongly and weakly overpressured regimes, which are
given by $M \propto \nu^{-1/3}$, will therefore correspond
also to a constant value of $L_{\rm j} / L_{\rm k}$. 
These
considerations tell us that the separation between the two regimes is
essentially determined by the jet kinetic power: the transition will
occur at higher values of $l_{\rm c}$ for high kinetic power jets and at lower $l_{\rm c}$ for low kinetic power jets.
For instance the dashed curve shown in Fig. 
\ref{mnuplane} representing the separation between the two regimes for
a cocoon length equal to 2 core radii 
corresponds to $L_{\rm j} / L_{\rm k} \sim 4.7 \times 10^3$.
The properties of the environment enter in the parameter $L_{\rm k}$ that fixes
a measure for the jet kinetic power.
For a typical cluster environment with
a central density $ 10^{-2}\, {\rm cm}^{-3}$, a core radius $100$ kpc 
and a temperature $3$ keV we have $L_{\rm k} = 6 \times 10^{42}\, {\rm erg\,  s}^{-1}$ 
and the dividing kinetic power for a cocoon length equal to 1.5 core radii
will be $L_{\rm j} \sim  1.9 \times 10^{46}\, {\rm erg\,  s}^{-1}$.
It is important to notice that for jets that are slightly
supersonic with respect to their internal sound speed the jet enthalpy flux
becomes comparable to the kinetic one. In these cases our estimates for
the jet kinetic power should be corrected at most by a factor two in order to
obtain an approximation of the total power of the jet.

Considering the particular case of Cygnus A, its cluster
environment is characterized by a central density $0.07\, {\rm cm}^{-3}$, 
a core radius $35$ kpc and a temperature $3.4$ keV
(Carilli et al. 1994), giving
$L_{\rm k} = 6.2 \times 10^{42}\, {\rm erg \, s}^{-1}$. The radio lobes of Cygnus A
show an extent of $\sim 70$ kpc, twice the core radius
($l_{\rm c}=2$). For this
length the dividing power is given by $L_{\rm j}/L_{\rm k} \sim 4.7 \times 10^3$,
as it is possible to see in Fig. \ref{mnuplane}. Since Cygnus A clearly
shows deficit of X-ray emission and therefore is in a weakly overpressured
regime, these estimates give an upper limit to the kinetic power of its jet
$\sim 2.9 \times 10^{46} \, {\rm erg \, s}^{-1}$. Nevertheless the pressure of the 
expanding cocoon must be higher than the ambient one since observations
by Chandra (Smith et al. 2002) show clearly that the shell is slightly
hotter than the surrounding medium. Then the cocoon is still expanding
as a (weak) shock wave more than a sound wave.

Another well known example of the interaction between a radio source and
the thermal gas of a cluster is the FR I type radio galaxy 3C~84 inside the 
Perseus cluster. The cluster core can be modeled with a central electron 
density $0.04 \, {\rm cm}^{-3}$, a core radius $a \sim 50$ kpc and a central 
temperature $T \sim 3.1$ keV (Schmidt, Fabian \& Sanders 2002) yielding
$L_{\rm k} = 6 \times 10^{42} \, {\rm erg\, s}^{-1}$. The radio lobes show an average extent of $\sim 
22$ kpc, giving a dividing power $L_{\rm j}/L_{\rm k} \sim 5.2 \times 10^2$. These
estimates give  an upper limit on the kinetic luminosity of the jet
$L_{\rm j} = 3.1 \times 10^{45} \, {\rm erg\, s}^{-1}$. This upper limit can be lowered  
considering that the shell of enhanced emission surrounding the cavities 
contains clearly the X-ray coolest gas in the cluster (Fabian et al. 2001). 
This fact rules out
the presence of strong shocks driven by the expanding cocoon. In our scheme
this means that the cocoon must be in a weakly overpressured regime in the
first phase of expansion yet as for example in the case $M=10$, $\nu=0.1$. 
As it can be seen in Fig. \ref{fig:tempdistr10} the average temperature of the shell for this
simulation is equal to the ambient one and in the shell there is also emission from gas cooler
than the ambient one. This cooling is due to the adiabatic expansion
of the shell after being compressed by a weak shock, in a way similar to that 
described in the simulations of expanding hot bubbles  
by Brighenti \& Mathews (2002). Taking this
case ($L_{\rm j}/L_{\rm k} = 100$) to determine a limit on the kinetic power of the jet 
we obtain $L_{\rm j} = 6.3 \times 10^{44} \, {\rm erg\, s}^{-1}$. 
For deriving the total jet power this estimate should be corrected by the enthalpy 
term, that for the above values of $M$ and $\nu$ is approximately equal to the $30\%$ of the kinetic power.
This estimate is similar to the
one by Fabian et al. (2001) who used the analytical bubble model by Churazov et al. (2000) to
determine their limits. This substantial agreement can be understood considering that
the Churazov et al. (2000) model solves exactly Eq. (\ref{eq:rci}) with spherical symmetry
 assuming that the bubble pressure is equal to the external one.
The limits of Fabian et al. (2001) are then determined requiring
that the bubble has the observed dimensions, that it expands subsonically 
and that it is not buoyant. This situation is similar to what we refer to as a weakly
or not overpressured regime.
The great difference between our low power cases
and a hot underdense bubble is the presence of the highly collimated jet that forms
a thermal hot spot and is likely not effective in uplifting 
the cooling flow gas as proposed for example by 
Soker et al. (2002) to explain the presence of cool gas in the rims around the cavities.
A collimated jet tends to go through the gas of the cluster displacing it aside and
without uplifting it. Once the jet has terminated its active phase is rapidly
destroyed (Reynolds, Heinz \& Begelman  2002) and the lobes can evolve like
buoyant bubbles (see for example Churazov et al. 2001 or Quilis, Bower \& Balogh 2001).
An example similar to Perseus A is the Hydra A cluster containing the powerful FR I radio 
source 3C~218. Given the properties of the X-Ray gas ($a \sim 26$ kpc, $n_0 \sim 0.06 \,
{\rm cm}^{-3}$, $T \sim 3.1$ keV, David et al. 2001) and a cocoon length of $\sim 49$ kpc
we obtain an upper limit on the jet power $L_{\rm j} \sim 1.1 \times 10^{46} \, {\rm erg\, s}^{-1}$.
Since
there is no indication that the gas surrounding the radio lobes is hotter than the ambient
cluster gas this limit can be lowered at least by an order of magnitude.
On the other hand it is possible to find some examples of radio sources inside galaxy clusters
that do not show the presence of X-ray cavities. In our scheme these sources could correspond
to radio lobes still in a strongly overpressured phase. For example the FR II radio source 3C~295
is found inside a cluster whose core is characterized by a radius $a \sim 17.8$ kpc, a central
density $n_0\sim 0.16\, {\rm cm}^{-3}$ and a temperature $T \sim 3.7$ keV
(Allen et al. 2001). This cluster does not show any cavity in its X-ray emission.
Given the longitudinal dimension of the radio source $\sim 17.6$ kpc we can estimate
a lower limit of the jet power $L_{\rm j} \sim 7.1 \times 10^{45} \,{\rm erg\, s}^{-1}$.

\section{Summary}
In this paper we performed numerical simulations of axisymmetric
supersonic jets propagating in an isothermal background atmosphere.
In agreement with the results presented by Reynolds, Heinz \& Begelman (2001), we
find two distinct and subsequent regimes of interaction between the
cocoon and the external medium. In the first phase of evolution, the
overpressured cocoon drives a strong shock in the ambient medium,
forming a thin, hot and compressed shell of shocked material, in the
second phase the shock becomes very weak and the shell widens,
decreasing its density and temperature.  The resulting X-ray
morphology in the two phases is different: in the strongly
overpressured phase, we expect a shell of enhanced X-ray emission
surrounding the radio emitting material, while, in the weak shock
phase, we expect a deficit of X-ray emission coincident with the radio
lobes. We have studied the dependence of the transition
between these two phases on the physical jet parameters, by a wide
coverage of the parameter space and by a comparison of the results of
numerical simulations with analytical models. We find that the
transition length between the two regimes depends essentially only on
the jet power scaled over a value dependent on the properties of the
ambient medium.

\begin{acknowledgements}
The authors acknowledge the italian MIUR for financial support, grant
No. 2001-028773. The numerical calculations have been performed at CINECA
in Bologna, Italy, thanks to the support of INAF.
\end{acknowledgements}

\appendix  
\section{}
 As we have discussed, the jet replenishes the cocoon
of matter and energy, therefore the cocoon expands compressing the
ambient medium and if the expansion speed of the cocoon is highly
supersonic, it will drive a strong shock in the external medium.
The first attempts to give an analytic description of theses processes
was done by Begelman \& Cioffi (1989) and Cioffi \&
Blondin (1992). To describe the energy input by the jet, they write
the following energy equation for the extended cocoon
\begin{equation}
  \label{eq:energy1}
  \frac {d E_{\rm c}}{dt} = L_{\rm j} \; ,
\end{equation}
where $E_{\rm c}$ is the total energy in the extended cocoon and $L_{\rm j}$
is the jet power. Assuming a constant energy input by
the jet and that all the energy is
converted in thermal energy, we can then write the average (extended) cocoon
pressure as
\begin{equation}
  \label{eq:pressure1}
  P_{\rm c} = \frac{(\gamma -1)L_{\rm j} t}{V_{\rm c}} \; , 
\end{equation}
where $V_{\rm c}$ is the volume of the (cylindrical) cocoon  equal to $\pi r^2_{\rm e}
l_{\rm c}$, $r_{\rm e}$ is the extended cocoon radius and $l_{\rm c}$  the cocoon length
	assumed to increase as $v_{\rm h} t$, with $v_{\rm h}$  given by the
one-dimensional estimate $v_{\rm h} = \sqrt{\nu} \ v_{\rm j} / (1+ \sqrt{\nu})$, where
$\nu = \rho_{\rm j}/\rho_{\rm ext}$.
The lateral expansion of the extended cocoon can then be obtained by assuming a
strong lateral shock, for which we can write
\begin{equation}
  \label{eq:rc_1}
  \frac{dr_{\rm e}}{dt} = \sqrt {\frac{\gamma+1}{2} \frac {P_{\rm c}}{\rho_{\rm ext}}} \; ,
\end{equation}
where $\rho_{\rm ext}$ is the external density. Equation (\ref{eq:rc_1}) can
be integrated giving for the time behavior of $r_{\rm e}$ a dependence on
$t^{1/2}$ and for $P_{\rm c}$ a dependence on $t^{-1}$. The behavior of the
cocoon proper can then be obtained by writing its energy balance, 
that will differ from Eq. (\ref{eq:energy1}) by the term describing the
work done at the contact discontinuity by the cocoon on the external
medium and will have the form 
\beq
\label{eq:rci}
\frac {dE_{\rm ci}}{dt} = L_{\rm j} - P_{\rm c} \frac {dV_{\rm ci}}{dt} \; ,
\eeq
where $E_{\rm ci}$ is now the total energy of the cocoon proper and
$V_{\rm ci}$ its volume. This equation can be written in an integral
form, neglecting the initial value of the volume:
\begin{equation}
V_{\rm ci} = \frac{1}{P_{\rm c}(t)^{1/\gamma}}\left\{ \frac{\gamma-1}{\gamma}
\int_0^t L_{\rm j} P_{\rm c}^{(1-\gamma)/\gamma} dt \right\} \; .
\label{eq:vci}
\end{equation}
Inserting in this equation the expression for
the pressure derived above and assuming that the volume $V_{\rm ci}$ is
proportional to $l_{\rm c} r_{\rm i}^2$ and the cocoon length behaves
in the same way as that of the extended cocoon we can derive that 
the cocoon radius $r_{\rm i}$ behaves as $t^{1/2}$, in the same way as
$r_{\rm e}$. In this situation we will then have a thin shell whose relative
thickness stays constant. This description is well suited for strongly
overpressured cocoons that drive strong shocks in the ambient medium,
but the temporal dependence of pressure ($\propto t^{-1}$) tells us
that a cocoon cannot stay during all its evolution in a strongly
overpressured regime. In addition there might be values of the
parameters for which the cocoon is not strongly overpressured already
at the beginning of its evolution. For describing these situations we
have to extend Begelman \& Cioffi (1989) model.  We need to reconsider
the energy balance of the extended cocoon, observing that in addition
to the energy input by the jet, the extended cocoon acquire also the thermal
energy of the external material that enters the extended cocoon during
its expansion. We can then write
\[
E_{\rm c} = L_{\rm j} t + \int_{V_{\rm c}} \frac {P_{\rm ext}} {\gamma - 1} dV \; .
\label{eq:energy2}
\]
where the second term on the right hand side represents the
contribution described above and $P_{\rm ext}$ is the pressure of the
external medium. The equation
for the pressure, instead of Eq. (\ref{eq:pressure1}) becomes
\begin{equation}
  \label{eq:pressure2}
  P_{\rm c} = P^* + \frac{(\gamma -1)L_{\rm j} t}{V_{\rm c}} \; ,
\end{equation}
where
\begin{equation}
  \label{eq:pstar}
  P^* = \frac{1}{V_{\rm c}} \int_{V_{\rm c}} P_{\rm ext} dV \; .
\end{equation}
In the case of uniform external pressure, we have, of course, that
$P^*=P_{\rm ext}$.
A second extension to the Begelman \& Cioffi (1989) model is done by
considering a lateral shock of arbitrary strength and writing the
expansion speed of the cocoon in a more general way as (see e.g. Landau \& Lifshitz, 1959)
\begin{equation}
\left(\frac{dr_{\rm e}}{dt}\right)^2 = c_{se}^2 \left( \frac{\gamma - 1}{2 \gamma} + \frac {\gamma +
    1}{2 \gamma } \frac {P_{\rm c}}{P_{\rm ext}} \right) \; .
\label{eq:vcocoon}
\end{equation}
The system of equations (\ref{eq:pressure2} - \ref{eq:vcocoon}) can be
solved analytically in the case of uniform external pressure assuming
a cocoon length $l_{\rm c} \sim \nu^{1/2} v_{\rm j} t$ and $L_{\rm j} = \pi/2 r_{\rm j}^2
\rho_{\rm j} v_{\rm j}^3$,i.e. the kinetic jet power. This approximation is
strictly valid for jets that are greatly supersonic with respect to their internal
sound speed, so as to neglect the enthalpy flux term.
Neglecting its initial value $ \sim r_{\rm j}$ (jet radius),
the cocoon radius then is given by
\beqar
\nonumber
 r_{\rm e}^2 \, & = & \, c_{se}^2 t^2 + \sqrt{\gamma^2-1} r_{\rm j} \nu^{1/4} M c_{se} t \, = \\
\nonumber    \\
\label{eq:rcocoon}
 & = &\, r_{\rm j}^2 \left( \frac{1}{\nu M^2} \frac{l_{\rm c}^2}{r_{\rm j}^2} + \frac{\sqrt{\gamma^2-1}}{\nu^{1/4}} 
  \frac{l_{\rm c}}{r_{\rm j}} \right)
\eeqar
and the cocoon pressure is

\beqar
\nonumber
\frac{P_{\rm c}}{P_{\rm ext}} \, & = & \, 1+\frac{\gamma(\gamma-1)}{2}
\left(\frac{1}{\nu^{1/2} M^2}\frac{c_{se}^2 t^2}{r_{\rm j}^2} 
+ \frac{\sqrt{\gamma^2-1}}{\nu^{1/4} M}\frac{c_{se} t}{r_{\rm j}}\right)^{-1} \, = \\
\nonumber \\
\label{eq:pcocoon}
\, & = &\, 1+\frac{\gamma(\gamma-1)}{2}
\left( \frac{1}{\nu^{3/2} M^4} \frac{l_{\rm c}^2}{r_{\rm j}^2} + \frac{\sqrt{\gamma^2-1}}{\nu^{3/4} M^2}
  \frac{l_{\rm c}}{r_{\rm j}} \right)^{-1} \, .
\eeqar
Substituting the solution of the pressure in the weakly overpressured regime in 
Eq. (\ref{eq:vci}), we obtain that the internal radius begins to expand subsonically
and tends asymptotically to a constant value.

\end{document}